# Polyimides Crosslinked by Aromatic Molecules and Nanocomposites for High-Temperature Capacitive Energy Storage

Feng Wang[1], Hao Wang[1], Xiaoming Shi[2], Chunli Diao[1], Chaolong Li[1], Weikun Li[1], Xu Liu[1], Haiwu Zheng[1, a)], Houbing Huang[2, a)], Xiaoguang Li[3]

[1] *Henan Province Engineering Research Center of Smart Micro-nano Sensing Technology and Application, School of Physics and Electronics, Henan University, Kaifeng 475004, P. R. China*

[2] *Advanced Research Institute of Multidisciplinary Science, and School of Materials Science and Engineering, Beijing Institute of Technology, Beijing 100081, P. R. China E-mail address: Houbing Huang: hbhuang@bit.edu.cn*

[3] *Hefei National Laboratory for Physical Sciences at the Microscale, Department of Physics, and CAS Key Laboratory of Strongly-coupled Quantum Matter Physics, University of Science and Technology of China, Hefei 230026, P. R. China*

[a)] Author to whom correspondence should be addressed: zhenghaiw@ustc.edu, hbhuang@bit.edu.cn.

**Abstract**

High-temperature polymer-based dielectric capacitors are crucial for application in electronic power systems. However, the storage performance of conventional dielectrics polymer dramatically deteriorates due to the thermal breakdown under concurrent high temperatures and electric fields, and there are hardly reports on the causes of thermal breakdown from the aspects of the high-temperature conduction loss and Joule heat dissipation. Herein, a combined strategy of crosslinking and compositing for polyimide-based nanocomposites is proposed, which minimizes the thermal breakdown by significantly inhibiting the high-temperature conduction loss and enhancing the high thermal conductivity. Furthermore, the rationale of the strategy was theoretically and experimentally verified from multiple perspectives. The charge-trapping effect is directly observed and quantitatively probed by Kelvin probe force microscopy with nano-level resolution, indicating that the crosslinking network introduces local deep traps and effectively suppresses the charge transport. The thermal conductivity of the nanocomposites inhibits the high-temperature thermal breakdown, which is confirmed by phase-field simulations. Consequently, the optimized nanocomposites possess an ultra-high discharge energy density ($U_d$) of 5.45 J/cm$^3$ and 3.54 J/cm$^3$ with a charge–discharge efficiency ($\eta$) of 80% at 150 and 200 ºC, respectively, which outperforms the reported polyimide-based dielectric nanocomposites. This work provides a scalable direction for high-temperature polymer-based capacitors with excellent performance.

Dielectric capacitors have played a key role in modern advanced electronics and circuit systems for controlling and storing charges owing to their intrinsic fast charge–discharge capability and extremely high power density.[1-3] Polymer-based dielectric materials with superior advantages, including ultra-high breakdown strength and outstanding reliability as well as excellent processability, are consistently preferred dielectrics for high-energy-density capacitors in comparison to inorganic dielectrics.[4-6] With the increasing calls for electric vehicles, underground oil/gas exploration, aerospace systems, and other emerging applications, there has been a new upsurge in the electric energy storage technology of polymer capacitors.[7, 8] These applications have placed higher demands on dielectric polymer stability and reliability when operating at high temperatures.[9-14] High-aroma polymers polyimides (PI) with a $T_g$ of 360 °C have been widely developed as high-temperature electronic applications and advanced composite materials owing to their exceptional thermal stability exceeding 100 ºC, electrical insulation performance and stable dielectric performance.[15-17] However, the conjugated segments in aromatic polymers promote charge transport at high electric fields, which leads to dramatically increased leakage current and thus the significant conduction loss.[10, 16] Such a reduced insulation performance of the dielectric polymer gives rise to not only a sharp fall in the charge–discharge efficiency ($\eta$) and energy storage density ($U_e$) during the charge–discharge cycles, but also a premature thermal breakdown of polymer films because of the thermal instability caused by the Joule heating.[18, 19] Crosslinking structures have been proven to effectively suppress conduction loss and ameliorate high-temperature capacitor capability by the migration of bound carriers.[20, 21] For example, Wang et al. tailored different crosslinking structures of dielectric polymers at the molecular level, and obtained an ultrahigh $U_e$ and $\eta$ at 150 °C.[22] Additionally, recent studies have demonstrated that introducing inorganic nanofillers with a wide bandgap into polymers induces a lower conduction loss and higher breakdown strength. This is currently the most effective method of fabricating high-temperature dielectric capacitors.[23-25] For example, Dai et al. introduced boron nitride nanosheets (BNNS) into polyimide-poly (amic acid) copolymer, and it was found that insulating BNNS was effective in impeding the leakage current and improving the high-temperature capacitive performances of the polymer. At 200 MV/m, the optimized $U_d$ reached 1.38 J/cm$^3$ with an $\eta$ value higher than 96% at 150 °C.[26] Although high-temperature dielectric polymer materials have been extensively studied, identifying a single strategy to address the multiple problems of high-temperature conduction loss and thermal breakdown of polymers at high temperatures is a huge challenge. It inevitably faces multi-step processing and high load, which is incompatible with the current methods used for large-scale fabrications of polymer dielectrics.

Compared to the above-mentioned single strategy, the combination strategy is anticipated to display more superiority. In this context, we propose a brand new and facile crosslinking structure, which was prepared by crosslinking P(I-AA) polymer with phenol-formaldehyde resin (PF). Taking a step forward, high thermal conductivity AlN nanoparticles with content ranging from 0 to 2 wt% were filled into the crosslinked polymers. In addition, while the energy storage performance of crosslinked polymer-based nanocomposites has been reported, the enhancement mechanism of the breakdown strength is limited to exploring the thermal breakdown caused by the increase of macroscopic leakage current at high temperatures, and there are almost no reports on the reasons of high-temperature thermal breakdown from the aspects of the conduction loss at high temperatures and Joule heat dissipation. Consequently, it is very necessary and highly desirable to deeply explore the binding of carriers by crosslinking networks, the dissipation of Joule heat, and the theoretical analysis of thermal breakdown suppression by high thermal conductivity materials at high temperatures. To be specific, the isothermal surface potential decay (ISPD) measured by Kelvin probe force microscopy (KPFM) of the crosslinked polymer reveals the trapping effect of the crosslinking structure on carriers. Moreover, the local trap level distribution calculated by the ISPD model shows that the crosslinking structure introduces charge traps and increases the trap depth. The embedding AlN nanoparticles with high thermal conductivity into the copolymer results in an enhanced λ of 16% at 150 ºC in comparison with that of the pure polymer (0.305 W/(m·K)). Such improvement in the heat dissipation performance of nanocomposites effectively enhances the breakdown performance at high temperatures. The suppression of thermal breakdown by the improvement of the high-temperature thermal conductivity of the composite is theoretically verified by phase-field simulation. This work represents an example of optimizing the molecular structure and compositing inorganic filler to create a high-temperature dielectric polymer, which results in a high $U_{\eta 80}$ of about 5.4 J/cm$^3$ at 150 °C. More impressively, when raising the temperature to 200 °C, the nanocomposite still exhibits an excellent capacitive performance.

The application scenarios and preparation process of nanocomposites are shown in Fig. 1(a) and 1(b), respectively. Firstly, PI-PAA copolymers with various components of PI were successfully prepared by controlling the processing temperature and heating time to regulate the thermal imidization, as shown in Fig. 1(b). From the chemical structure of PAA and PI, dehydration cyclization (2H$_2$O per replicating unit) takes place during the acylation process, and –COOH in PAA transforms to an imide ring in PI.[27] Note that –COOH is the polar group, and the imide ring is rigid, limiting the movement of the chains. Fourier transform infrared spectroscopy (FTIR) was performed to determine the composition of the PI-PAA copolymer, as shown in Fig. S1(a) and (b). The typical unipolar *D*–*E* loops of copolymers with various PI content at 25, 120, and 150 °C are drawn in Fig. S2(a)-S2(c). Meanwhile, the $U_{\eta 80}$ values of the copolymers at different temperatures calculated by the *D*–*E* loops are shown in Fig. S2(d). The results indicate that with the increasing temperature, the $U_{\eta80}$ of the 0.88PI-0.12PAA copolymer is better than that of other components, which is related to the incremental effect of the polar group –COOH for $U_d$ at high temperatures. For simplicity, the 0.88PI-0.12PAA is labelled as P(I-AA). On the basis of the above-mentioned copolymer, we presented a crosslinking structure that was constructed by blending P(I-AA) with PF. Subsequently, the mixed solutions were heated to form crosslinked polymer, which was labelled as P(I-AA-F). The reaction between the copolymer and –OH of PF forms –C–O–C linked with their respective main chains.[27] The crosslinking structure is confirmed by FTIR (Fig. S3). Taking a step forward, nanocomposites with crosslinked polymers and an AlN nanoparticles content ranging from 0 to 2 wt% were fabricated. The characterization of AlN nanoparticles is shown in Fig. S4. The surface SEM images of nanocomposites with various mass fractions of AlN nanoparticles are shown in Fig. S5(a)-S5(e). It can be seen that when the nanofiller content is relatively low, AlN nanoparticles are uniformly dispersed in the P(I-AA-F) matrix, and when the loading mass fraction reaches 2 wt%, an agglomeration of nanoparticles appears. The cross-sectional morphology of the nanocomposite with 1.5 wt% AlN is shown in Fig. S5(f), indicating the thickness is about 9 μm and no noticeable deficiencies were observed.

To simulate the transport rates of charges in the polymer dielectrics, we used KPFM with nano-level resolution to investigate the surface potential decay of P(I-AA) and P(I-AA-F) to obtain direct evidence of the capture traps introduced by the crosslinking structure.[29] Firstly, extrinsic electrons were injected into the middle parts (0.8×0.8 μm$^2$) which are in the target area (3×3 μm$^2$) of the samples, which were charged by scanning the middle section with the probe tip applied a DC voltage of 1.5 V. Then, the DC voltage was removed after the carrier injection, and the dynamic evolution of potential distribution on the surface of the target region was measured with an interval of 8 min.[30] The dependence of surface potential evolution of the P(I-AA) and P(I-AA-F) on time at room temperature is presented in Fig. 2(a), where it can be seen that the recovery of the target-region potential of P(I-AA) and P(I-AA-F) takes a long time, indicating that there are abundant trap points inside the film. Since the potential recovery of target region in P(I-AA-F) falls significantly behind that in P(I-AA), the electron trapping effect of the crosslinked polymer rather than the copolymer contributes more to the inhibition of hopping conduction. These results demonstrate that constructing a crosslinking structure can greatly enhance the charge-trapping depth and effectively suppress the migration of carriers. The horizontal distribution of local charge traps can be quantified using the ISPD model in the KPFM research[31] and Fig. 2(c) plots the calculated local trap level distribution. It is evident that the construction of the P(I-AA-F) with crosslinking structure not only introduces more charge traps but also increases the charge trap depth compared with the copolymer of P(I-AA). Apart from room temperature, the surface potential decay of P(I-AA) and P(I-AA-F) using the KPFM method was also investigated at 150 ºC. The potential decay maps for both P(I-AA) and P(I-AA-F) in the target area after scanning/charging are revealed in Fig. 2(b). Noticeably, the decay of charges in the P(I-AA-F) is much slower than that in the P(I-AA), which suggests that the construction of the crosslinking structure is still highly effective for restricting the charge transport and suppressing the conduction loss at high temperature.[32] Furthermore, hopping conduction is a kind of tunneling effect in which electrons trapped in dielectric polymers transfer from one trap site to another by hopping.[18] A deeply understanding of the inhibitory effect of the crosslinking structure on the conduction loss can be achieved from the schematic diagram, as shown in Fig. 2(d), which vividly illustrates the trapping effect of more localized deep traps on carriers, limiting the movement of carriers and thus reducing the leakage current.[33]

Then the dielectric properties and energy storage performance of all-organic polymer dielectrics were tested at room and high temperatures, as shown in Fig. S6 and S7. The crosslinking network limits dipole motions, which causes an obstacle to the conduction loss and effectively reduces the dielectric loss (tanδ).[28] It can be seen that the $U_d$ elevates with the increasing electric field at 150 °C, and the P(I-AA-F) exhibited excellent energy storage performances with $U_{\eta 80}$ of 2.88 J/cm³ compared to the P(I-AA) with $U_{\eta 80}$ of 0.89 J/cm³.

The addition of AlN as the insulating coating to dielectric polymer has been reported to effectively enhance barrier height of electron injection and to remarkably suppress conduction loss at high temperatures, improving the energy storage performance of nanocomposites.[34, 35] Fig. 3(a) and 3(b) exhibit the Weibull distribution of nanocomposites with different mass fractions of AlN nanoparticles at room temperature and high temperature, respectively. At room temperature, the $E_b$ of the nanocomposites increases from 618 MV/m to 661 MV/m with the addition of AlN nanoparticles, reaching the maximum when the loading fraction is 0.5 wt%. At a high temperature of 150 °C, the $E_b$ of the nanocomposites increases from 499 MV/m to 617 MV/m with the addition of AlN nanoparticles, reaching the maximum when the loading fraction is 1.5 wt%. Notably, $E_b$ significantly reduced when the mass fraction of AlN nanoparticles exceeds 1.5%, which is attributed to the agglomeration of the AlN fillers in the nanocomposites.[23] Excessive AlN fillers tend to cause more structural defects, reducing the $E_b$.[36, 37] To more clearly observe the evolution of $E_b$ at room temperature and high temperature, the $E_b$ of the nanocomposites as a function of mass fractions of AlN nanoparticles is summarized in Fig. 3(c). Since the benefit of thermal management of the nanocomposites lies in suppressing thermal breakdown, the enhancement peak of $E_b$ significantly varies from the composite AlN content at room temperature and high temperature. At high temperatures, thermal breakdown is the key mechanism for the breakdown of PI-based polymers.[18] Compositing AlN nanoparticles with high thermal conductivity is expected to increase the thermal conductivity of the nanocomposites. This is exactly what the high-temperature dielectric capacitor required. The thermal conductivity of the pure P(I-AA-F) and P(I-AA-F) based nanocomposite with 1.5 wt% AlN measured at three different temperatures was shown in Fig. 3(d), from which the incorporation of AlN into P(I-AA-F) indeed increase the thermal conductivity ($\lambda$) and the $\lambda$ of both P(I-AA-F) and the nanocomposite rises with the increasing temperature. The increasing rate of $\lambda(\Delta\lambda)$ can be expressed by the difference between the $\lambda$ of P(I-AA-F) and that of the nanocomposite at the same temperature. It is worth noting that the $\Delta\lambda$ for the two materials is different. For instance, $\Delta\lambda$ is 0.033 W/(m·K) at room temperature, while $\Delta\lambda$ is 0.047 W/(m·K) and 0.051 W/(m·K) at 120 ºC and 150 ºC, respectively. These experimental results corroborate that compositing AlN nanoparticles into P(I-AA-F) matrix can effectively improve the thermal conductivity of the nanocomposite. The fact that the improvement in thermal conductivity is particularly evident at high temperatures signifies the introduction of AlN nanoparticles into polymers is an effective and facile strategy for improving dielectric capacitors operated at high temperatures.

To further explore the role of the AlN nanoparticles in the enhancement of $E_b$ at different temperatures, phase-field simulations were employed to investigate the breakdown phase evolution of P(I-AA)-based composite structures.[38] Here in our simulation, the avalanche-type electronic breakdown was considered as the dielectric breakdown mechanism in the nanocomposites.[39] An electric field along the $z$-direction was applied to drive the breakdown phase evolution. Due to the ultra-wide band gap of AlN nanoparticles, a higher electric field exists in the AlN nanoparticles, and the high $E_b$ of AlN can block the evolution of the broken domain. Next, we adopted different mass fractions of AlN nanocomposites to investigate the breakdown evolution at room temperature and high temperature, as shown in Fig. 4. The $E_b$ increases firstly (0–0.5 wt% AlN) but then decreases (>1%) at room temperature. At 150 ºC, the decrease of $E_b$ occurs at a higher volume fraction of AlN. That is consistent with the above-mentioned experimental conclusion. The reason is that the breakdown probability at high temperatures is higher than that at room temperature, resulting in a fast breakdown velocity. The breakdown path can quickly bypass the fewer AlN nanoparticles until the number of AlN nanoparticles is large. The higher mass fraction of AlN nanoparticles, the larger thermal conductivity of nanocomposites, which can effectively reduce the effect of thermal breakdown on the energy storage performance of the current dielectric polymer. In this case, it is not difficult to understand that the AlN content for the maximum $E_b$ at 150 ºC is 1.5 wt% rather than 0.5 wt%. Fig. S8 presents dielectric properties and the leakage densities of the nanocomposites with different nanoparticles content at room temperature, from which it seems that introducing an appropriate amount of AlN nanoparticles can effectively improve the insulation performance and maintain stable dielectric properties over a wide temperature range from room temperature to 200 ºC of the nanocomposite. Briefly, the phase-field simulation shows that the breakdown path in nanocomposite can be effectively suppressed by the highly loaded AlN nanoparticles at high temperatures, leading to an enhancement of $E_b$.

Representative unipolar D–E loops of the nanocomposites at different temperatures are plotted in Fig. S9(a)-S9(c), respectively. Although $D_{max}$ decreases with the increasing mass fraction of AlN nanoparticles under the same electric field, sacrificing polarization to improve the $E_b$ and $\lambda$ is more beneficial for the energy storage of the nanocomposite. Meanwhile, the $U_{\eta 80}$ values of the nanocomposites at different temperatures calculated by the *D–E* loops are shown in Fig. S9(d). Fig. S10 and Fig. 5(a) plot the $U_d$ of the nanocomposites at room temperature and high temperature with the increasing electric field, and the nanocomposite with 1.5 wt% AlN achieves a high $U_{\eta 80}$ of 5.45 J/cm$^3$ at 150 ºC. To further research the energy storage performance of the nanocomposites at higher temperatures, the unipolar *D–E* loops of the nanocomposite with 1.5 wt% AlN at different electric fields were tested at 200 ºC, as shown in Fig. 5(b). It is gratifying that the excellent $U_d$ of 3.90 J/cm$^3$ and $\eta$ of 76% are still exhibited under 530 MV/m even at high temperature of 200 ºC. Fig. 5(c) summarizes the maximum $U_d$ achieved by the current dielectric polymers and the nanocomposite with the $\eta$ exceeding 80% at 150 °C. Notably, the nanocomposite prepared in this work obtained an excellent $U_{\eta 80}$ through a simple and low-cost processing technology, surpassing most of reported dielectric capacitors. As shown in Fig. 5(d), the maximum $U_d$ of the nanocomposite shows superiority compared to the publicly reported dielectric capacitors operated at 200 °C. The cycling reliability and discharge behavior of the nanocomposites are shown in Fig S11 and S12. The results show that the AlN with high thermal conductivity has broad prospects for reducing waste heat and increasing the operating temperature of dielectric capacitors based on polymers.

In summary, we have successfully substantiated that polyimide-based nanocomposite with a crosslinking structure exhibits a significantly enhanced high-temperature capacitance performance, outperforming the currently reported dielectric polymers with a single strategy. More importantly, we have deeply explored the rationality of the strategy from multiple aspects and perspectives through experimental results and computational simulation. The binding effect of crosslinked networks on carriers was directly observed by KPFM with nano-level resolution. The suppression of high-temperature thermal breakdown is theoretically verified by phase-field simulation, demonstrating that the introduction of AlN nanoparticles is an effective and facile strategy to improve dielectric capacitors operating at high temperatures. The optimized nanocomposite possesses reduced high-temperature conduction loss and low probability of thermal breakdown, which give rise to the significant improvement in $U_d$ of 3.90 J/cm$^3$ at 200 ºC. Meanwhile, excellent charge–discharge cycling stability and ultrafast discharge capacity improve the competitiveness of the nanocomposite in dielectric capacitors operating at high temperatures. We anticipate that the well-designed strategy of nanocomposites will provide broad prospects for the research of high-energy-density dielectric polymers operating at high temperatures.

**AUTHOR'S CONTRIBUTIONS**

**Feng Wang** and **Hao Wang** contributed equally to this work.

**Feng Wang:** Data curation (lead); Formal analysis (lead); Writing original draft (equal). **Hao Wang:** Data curation (supporting); Formal analysis (supporting); Writing original draft (equal). **Xiaoming Shi:** Formal analysis (supporting). **Chunli Diao:** Data curation (supporting); Formal analysis (supporting). **Chaolong Li:** Data curation (supporting); Formal analysis (supporting). **Weikun Li:** Data curation (supporting); Formal analysis (supporting). **Xu Liu:** Data curation (supporting); Formal analysis (supporting). **Haiwu Zheng:** Conceptualization (lead); Project administration (lead); Methodology (lead); Writing – review & editing (lead). **Houbing Huang:** supervision (equal). **Xiaoguang Li:** Conceptualization (lead); Writing – review & editing (lead).

**ACKNOWLEDGMENTS**

The authors gratefully acknowledge the support from the National Natural Science Foundation of China (Nos. 52072111) and Natural Science Foundation of Henan Province in China (Nos. 212300410004).

**AIP PUBLISHING DATA SHARING POLICY**

The data that support the findings of this study are available from the corresponding author upon reasonable request.

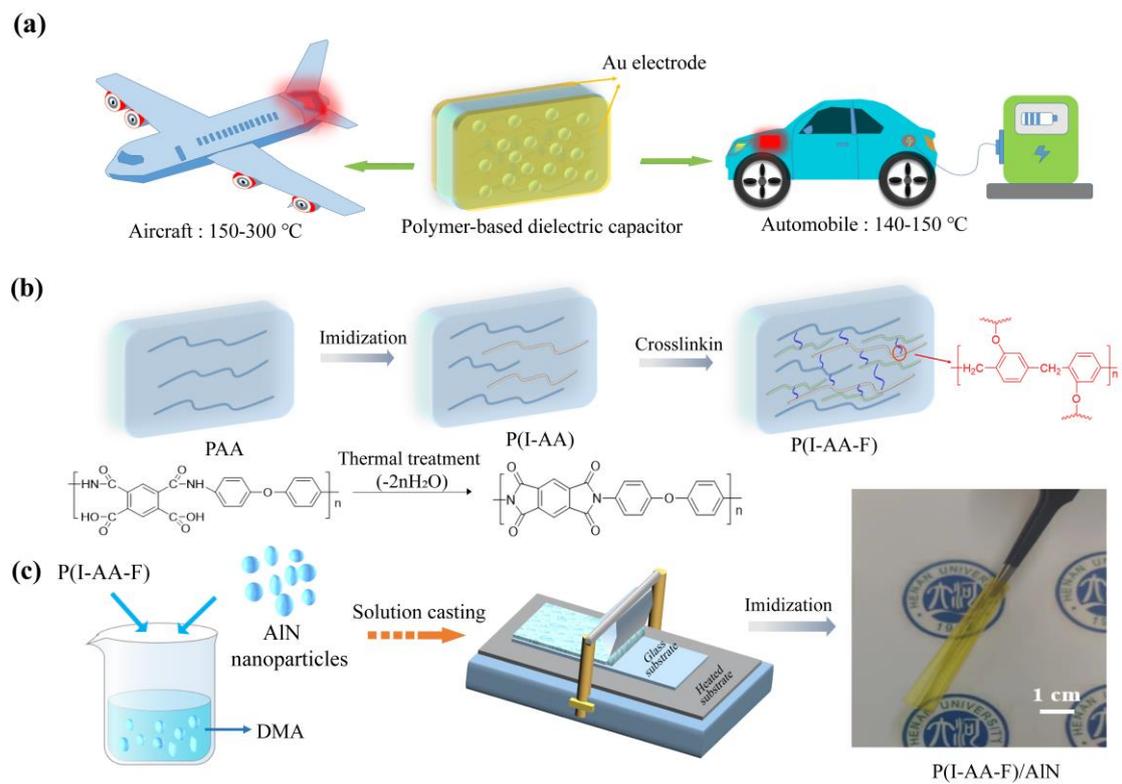

**Fig. 1**

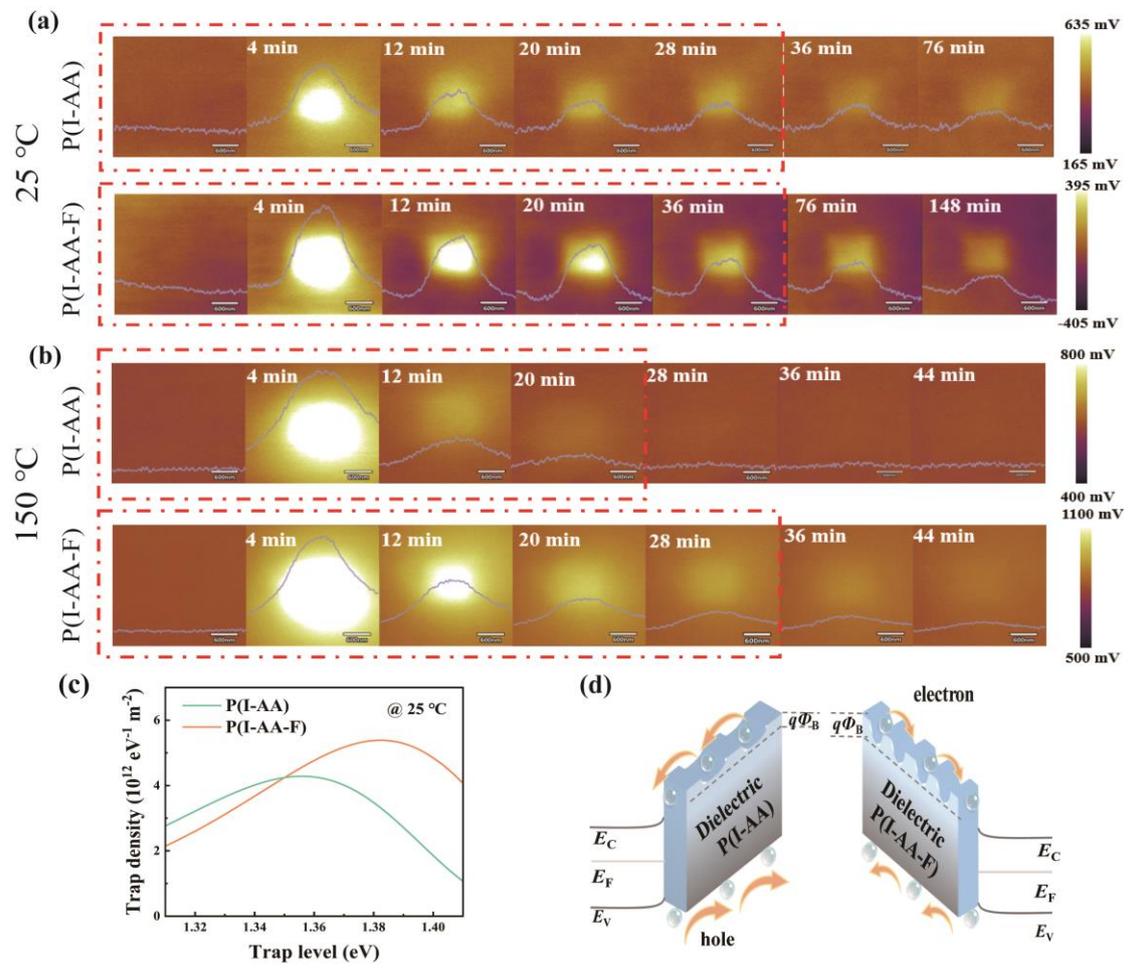

**Fig. 2**

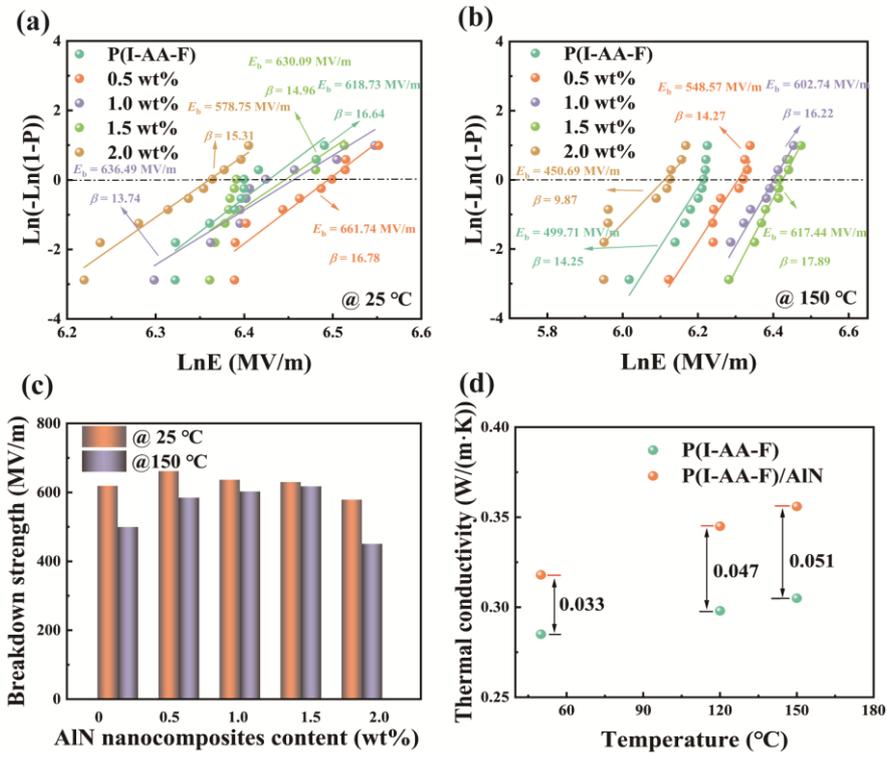

**Fig. 3**

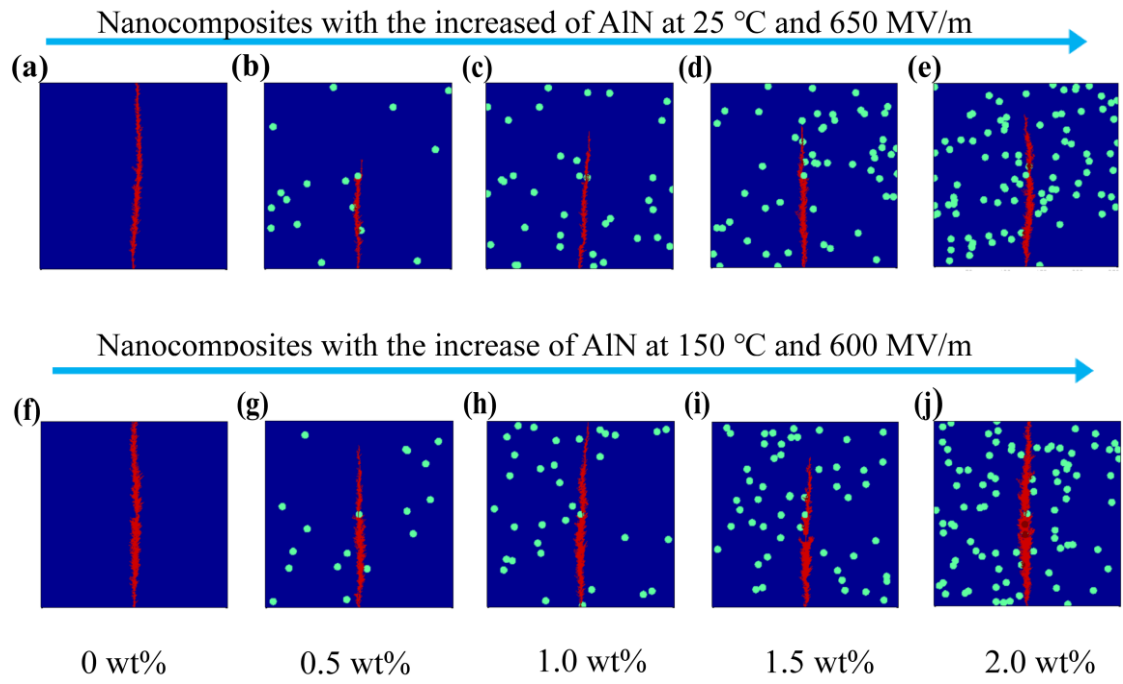

**Fig. 4**

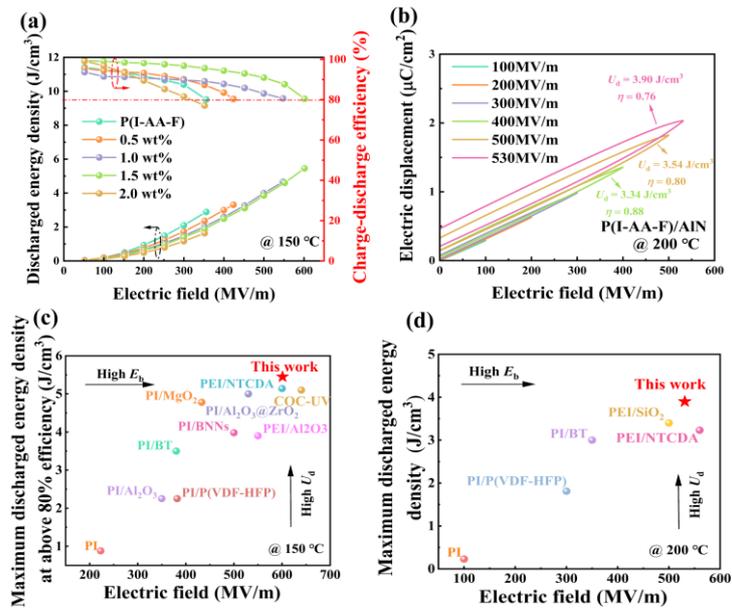

**Fig. 5**

**Fig. 1.** The application scenarios and preparation process of the nanocomposites. (a) Application scenarios of polymer capacitors. (b) Schematic and structures of copolymers and crosslinked polymer. (c) Diagrammatic sketch of the fabrication process of the P(I-AA-F) nanocomposites.

**Fig. 2.** The local charge trap level distribution characterized by KPFM. The images of P(I-AA) and P(I-AA-F) after injecting electrons by negative biasing at (a) 25 °C and (b) 150 °C, and the inset in each image indicates the potential profile along the light blue solid line. (c) Local charge trap level distribution at the interfacial region at 25 °C. (d) Hopping conduction in dielectric polymers. ($q\Phi_B$ is the Barrier height, $E_C$ is the Conduction band, $E_V$ is the Valence band and $E_F$ is the Fermi level).

**Fig. 3.** Weibull statistics of breakdown strength. Weibull distribution of nanocomposites with various AlN contents at (a) 25 °C and (b) 150 °C. (c) Comparison of the breakdown strength at 25 °C and 150 °C. (d) Thermal conductivity of crosslinked polymer and nanocomposite with 1.5 wt% AlN at three different temperatures.

**Fig. 4.** Electric field distributions and breakdown paths calculated by phase-field simulations. (a)–(e) The break evolution at 25 °C and 650 MV/m. (f)–(j) The break evolution at 150 °C and 600 MV/m of the nanocomposites with different mass fractions of AlN.

**Fig 5.** Energy storage performance of the nanocomposites. (a) $U_d$ and $\eta$ of the nanocomposites with various AlN nanoparticles contents at 150 °C. (b) $D-E$ loops of the nanocomposite with 1.5 wt% AlN at 200 °C. (c), (d) Comparisons of $E_b$ and the corresponding $U_d$ for previously reported results at 150 °C and 200 °C.